\documentclass[11pt,amsmath,amssymb,twoside,floatfix,showpacs,pre,superscriptaddress]{revtex4}

\usepackage{hyperref}
\usepackage[latin1]{inputenc}
\usepackage[english]{babel}
\usepackage{graphicx}
\usepackage[pdftex]{epsfig}
\usepackage{amsmath}
\usepackage{amssymb}
\usepackage{bm}
\usepackage{natbib}

\newcommand{\eq}[1]{Eq.\,(\ref{#1})}
\newcommand{\eqns}[2]{Eqns.\,\mbox{(\ref{#1})--(\ref{#2})}}
\renewcommand{\vec}[1]{\mathbf{#1}}
\newcommand{\fig}[1]{Fig.\,\ref{#1}}
\newcommand{\tab}[1]{Tab.\,\ref{#1}}
\newcommand{\ba}[1]{\begin{array}{#1}}
\newcommand{\ea}{\end{array}}
\newcommand{\etal}{~{\it et\,\,al.\,}}
\newlength{\figwidth}

\graphicspath{{}{images/}}
\renewcommand{\theta}{\vartheta}

\begin{document}
\setlength{\figwidth}{0.7\linewidth}

\title{A boundary condition with adjustable slip length for Lattice Boltzmann simulations}
\author{Nayaz Khalid Ahmed} 
\affiliation{National Institute of Technology\\ Tiruchirappalli, India 620 015} 
\author{Martin Hecht} 
\affiliation{Institute for Computational Physics, \\ Pfaffenwaldring 27, 70569 Stuttgart, Germany}

\date{\today}

\begin{abstract}
\noindent
\textbf{Abstract: }
A velocity boundary condition for the Lattice Boltzmann simulation technique has been proposed 
recently by Hecht and Harting\,\cite{Hecht2009}. This boundary condition is independent of the 
relaxation process during collision and contains no artificial slip. In this work, this boundary 
condition is extended to simulate slip-flows. The extended boundary condition has been tested 
and it is found that the slip 
length is independent of the shear rate and the density, and proportional to the BGK relaxation time. 
The method is used to study slip in Poiseuille flow and in linear shear flow. Patterned walls with 
stripes of different slip parameters are also studied, and an anisotropy of the slip length in 
accordance with the surface pattern is found. The angle dependence of the simulation results 
perfectly agrees with theoretical expectations. The results confirm that the proposed boundary 
conditions can be used for simulating slip-flows in micro fluidics using single relaxation time 
lattice Boltzmann, without any numerical slip, giving an accuracy of the second order.
\end{abstract}


\maketitle

\section{Introduction}
 
Micron- and submicron-size mechanical and biomechanical devices are becoming more prevalent both in commercial applications and in scientific inquiry\,\cite{Beskok05}. Small accelerometers with dimensions measured in microns are being used to deploy air bag systems in automobiles. Novel bioassay consisting of microfluidic networks are designed for patterned drug delivery. Inherent in these technologies is the need to develop the fundamental science of small devices. 

At the macroscale, where a continuum flow of a Newtonian fluid is assumed, molecular effects are integrated out. The dynamics of the averaged quantities is described by the Navier-Stokes equation, the governing equation of traditional fluid mechanics. Interactions among fluid molecules are expressed by material constants such as compressibility, and shear and bulk viscosities. The no-slip condition represents the interactions between the fluid and a solid surface. Both, viscosity and no-slip condition, are concepts developed under the framework of the continuum hypothesis. In micro- and nano-flows, the continuum assumption, or parts of it, break down, so that deviations from the  viscosity in the bulk and from the no-slip condition may occur.

The no-slip boundary condition implies that the velocity of the fluid flow tangential to the surface at the solid-fluid interface equals the velocity of the solid surface in the same direction. The no-slip boundary condition is at the centre of our understanding of fluid mechanics at the macro scale. As this condition cannot be derived from first principles, it could be violated, and this has been found to occur in certain cases of microflows\,\cite{lauga-brenner-stone}.

\begin{figure}[t]
\centering
\includegraphics[height=0.25\textheight]{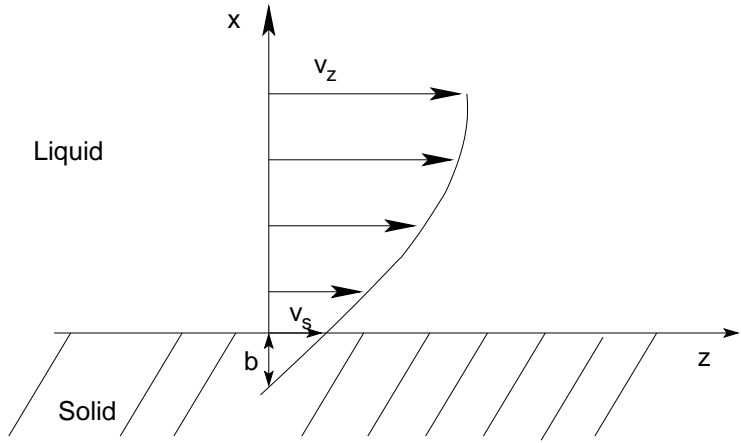}
\caption{Definition of the slip length\,\cite{Navier00}. The relative velocity of the fluid at the surface to the surface is denoted by $v_s$ and is called the slip velocity. The imaginary distance below the surface at which the extrapolated relative velocity of the fluid is zero, denoted by \emph{b}, is called the slip length.}
\label{fig:slip_length}
\end{figure}
The physical picture of slip that emerges is that of a complex behavior at a solid-fluid interface, involving interplay of many physicochemical parameters including wetting, surface charge, surface roughness, impurities and dissolved gases. Slip has been found to predominantly occur at hydrophobic surfaces, as well as in flows of non-Newtonian fluids such as polymer solutions. The slip flow of a fluid over a solid surface is quantified by the \emph{slip length}. The slip length is defined as the imaginary distance behind the solid-fluid interface at which the fluid velocity extrapolates to 
zero\,\cite{Navier00}, as shown in \fig{fig:slip_length}. The concept of slip was first introduced by Navier in 1824\,\cite{Navier00}. However, a century of agreement between experimental results in liquids and theories derived assuming the no-slip boundary condition resulted in further inquiry into the slip condition only with the advent of micro flows.\\
For nanoscale flows, slip is of technological utility. Small size of devices requires pumping liquids through nanochannels in processes such as desalination and other chemical purification techniques. However, if the liquid can be made to slip, then the resistance and the energy requirements can be reduced, with the promise that these techniques become economically viable\,\cite{lauga-brenner-stone}.

As a result of the breakdown of the continuum assumption in microflows, the classical Navier-Stokes equations cannot be applied to understand the fluid dynamics. Recent computer simulations apply molecular dynamics\,\cite{Rapaport95}. These simulations are usually limited to some tens of thousands of particles, length scales of nanometres and time scales of nanoseconds. Also, shear rates are usually orders of magnitude higher than in any experiment\,\cite{lauga-brenner-stone}. Due to the small accessible time and length scales of molecular dynamic simulations, mesoscopic simulation methods such as the Lattice Boltzmann method (LBM) are more applicable for the simulation of microfluidic experiments. The LBM solves the lattice Boltzmann equation (LBE), which is a discrete approximation to the continuous Boltzmann equation and has been recognized as a promising approach for simulation of microflows\,\cite{chen-doolen98}. It has been shown that in the macroscopic limit, the Navier Stokes equations can be recovered\,\cite{chen-chen-matthaeus,Higuera89}. However, most previous LBE models virtually correspond to the Navier-Stokes equations already on the length scale of the lattice constant, and when these models are applied to near-continuum flows, possible boundary effects have to be incorporated into the boundary conditions of the simulation. This way the influence of a possible partial slip on the 
flow can be studied. Thus the development of slip-flow boundary conditions is an important issue for the simulation of fluid flows with the LBE method.
We assume that the lattice Boltzmann method is operated in the near-continuum limit.

In this paper, we consider a recently proposed velocity boundary condition for the LBE that is independent of the relaxation process during collision and contains no artificial 
slip\,\cite{Hecht2009}. We extend the applicability of this boundary condition to 
simulate slip by the use of a parameter that defines the surface tendency to cause slip.
We make no assumptions for the origin of the slip. It may be caused by the surface
properties, i.e. flow of an aqueous solution over a hydrophobic surface, or flow of a 
rarefied gas so that a Knudsen layer causes the slip. For the latter case however, there exist more elaborated boundary conditions based on a diffusively reflecting wall\,\cite{Ansumali02}, which has been used in a number of works on gas micro-flows\,\cite{Sofonea05,Toschi05,Kim08,Zheng08}. The diffusively reflecting 
boundary condition\cite{Ansumali02} is more predictive for slip due to Knudsen-layer effects because it contains no artificial parameter, whereas we do not aim for resolving the Knudsen layer and assume that our approach phenomenologically captures slip flow of any origin, but leaves the calibration of the slip parameter as a separate task.

In the following section we shortly introduce the lattice Boltzmann method and then describe
the boundary conditions we use. In the section thereafter we present the results, first for 
Poiseuille flow between two parallel plates, then for linear shear flow, before turning to 
patterned surfaces, where the walls are textured with stripes of alternating slip. In the last 
concluding section we summarize the results.

\section{Simulation Method}
\label{sec_method}

\subsection{Lattice Boltzmann Method} The lattice Boltzmann method (LBM) is a numerical method to solve the Boltzmann equation\, \eq{eq_boltzmann} on a discrete lattice\,\cite{chen-doolen98}. The Boltzmann equation describes the dynamics of a gas from a microscopic point of view: in a gas, particles, each with velocities $\vec{v}_i$, collide with a certain probability and exchange momentum among each other.
For ideal collisions total momentum and energy are conserved in the collisions.
The Boltzmann equation expresses how the probability $f(\vec{x},\vec{v},t)$
of finding a particle with velocity $\vec{v}$ at a position $\vec{x}$ and at 
time $t$ evolves with time:
\begin{equation}
 \label{eq_boltzmann}
  \frac{{\rm d}f}{{\rm d}t} =
  \vec{v}\cdot\nabla_\vec{x}f + 
  \vec{F}\cdot\nabla_\vec{p}f +
  \frac{\partial f}{\partial t} =
  \hat\Omega(f)\,,
\end{equation}
where $\vec{F}$ denotes an external body force, $\nabla_{\vec{x},\vec{p}}$ the
gradient in position and momentum space, and 
$\hat\Omega(f)$ denotes the so-called collision-operator. Bhatnagar, Gross, 
and Krook proposed the so-called BGK dynamics\,\cite{BGK}, 
where the collision operator $\hat\Omega$ is chosen as a relaxation with a characteristic time $\tau$ to the equilibrium distribution $f^{(eq)}(\vec{v}, \rho)$.
\begin{equation}
   \label{eq_bgk}
   \hat\Omega(f) =-\frac{1}{\tau}\left(f-f^{(eq)}\right)\,.
\end{equation}

\begin{figure}[t]
\centering
\includegraphics[width=\figwidth]{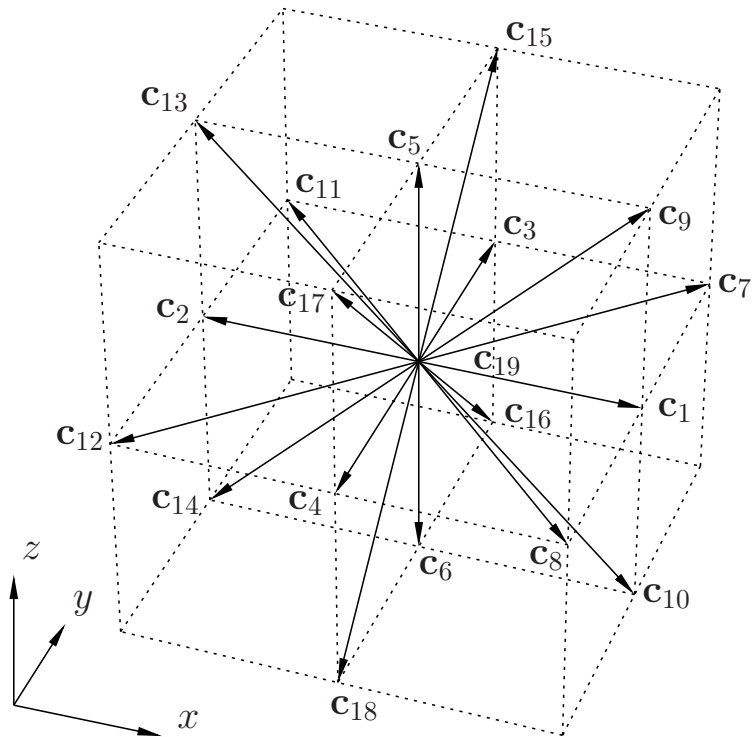}
\caption{The geometry of our D3Q19 lattice with the lattice vectors $\vec{c}_i$}
\label{fig_d3q19}
\end{figure}
When \eq{eq_boltzmann} is solved on a lattice, the velocity space is discretized. 
Different schemes to discretize the velocity space have been proposed, among which
the higher order schemes\cite{Chikatamarla06} might reproduce more features of the 
analytical Boltzmann equation than other schemes. However, if aiming for linear effects
or near-continuum flows, lower order schemes are sufficient. A widely used 3-dimensional 
lattice type is the D3Q19 lattice\,\cite{qian}, illustrated in \fig{fig_d3q19}, which
we use in this study.
Each node is joined to its neighbours by a set of lattice vectors $\vec{c}_i$. We use the notation that the vectors $\vec{c}_i$ are the $i^{th}$ column 
vector of the matrix 
\begin{widetext}
\begin{equation}
 \label{eq_deffs}
  \mathbf{M}=\left[\,
  \ba{ccccccccccccccccccc}
    \,1\,&\,-1\,&\, 0\,&\, 0\,&\, 0\,&\, 0\,&\, 1\,&\, 1\,&\, 1\,&\, 1\,&\,-1\,&\,-1\,&\,-1\,&\,-1\,&\, 0\,&\, 0\,&\, 0\,&\, 0\,&\, 0\, \\    
    \,0\,&\, 0\,&\, 1\,&\,-1\,&\, 0\,&\, 0\,&\, 1\,&\,-1\,&\, 0\,&\, 0\,&\, 1\,&\,-1\,&\, 0\,&\, 0\,&\, 1\,&\, 1\,&\,-1\,&\,-1\,&\, 0\, \\
    \,0\,&\, 0\,&\, 0\,&\, 0\,&\, 1\,&\,-1\,&\, 0\,&\, 0\,&\, 1\,&\,-1\,&\, 0\,&\, 0\,&\, 1\,&\,-1\,&\, 1\,&\,-1\,&\, 1\,&\,-1\,&\, 0\, \\
  \ea\,
  \right]\,.
\end{equation}
\end{widetext}

On such a discrete lattice, the local density at a lattice point can be obtained by
summing up all $f_i$,
\begin{equation}
  \rho(\vec{x}, t) = \sum\limits_{i=1}^{19}f_i(\vec{x},t)\,,
  \label{eq_rho}
\end{equation}
and the streaming velocity is given by
\begin{equation}
  \vec{v}(\vec{x},t) = \frac{1}{\rho(\vec{x},t)} \sum\limits_{i=1}^{19}f_i(\vec{x},t)\vec{c}_i\,.
  \label{eq_vel}
\end{equation}
Note that we express all quantities in lattice units, i.e., time is measured 
in units of update intervals and length is measured in units of the lattice constant. 

\subsection{Boundary Conditions}

On the wall boundary nodes, the distribution function assigned to vectors $\vec{c}_i$ pointing out of the lattice move out of the computational domain in the propagation step, and the ones assigned to the opposing vectors are undetermined because there are no nodes which the distributions could come from. Therefore, on the boundary nodes, special rules have to be applied.

Generally, in applications involving slip flows of a real fluid close to a wall, the structure of the fluid changes due to the interaction with the wall. This can be reflected in a modified ordering in the fluid, a modified mean free path for rarefied gases, or an electrostatic repulsion or Van der Waals attraction. 

The lattice analogue of this wall boundary condition is a blend of no-slip bounce back and full-slip specular reflections. To quantify this phenomena, we introduce a parameter $\zeta$ that behaves as a weighting function to promote slip.

\paragraph{Combined Boundary Condition}
In this work, we test a linear combination of two boundary conditions to model slip flows through a slip parameter $\zeta$ by the following general relation:
\begin{equation}
\label{eq_slip_param}
f_i^{ps} = \zeta \* f_i^{fs} + (1-\zeta) \* f_i^{ns} \,,
\end{equation}
where $f_i^{ps}$ denotes the density probability function for flow with the partial-slip condition, $f_i^{ns}$ denotes the density probability function of a boundary condition which generates the no-slip condition, and $f_i^{fs}$ stands for a boundary condition from which full-slip condition may be obtained. This approach is widely used\,\cite{Succi01,Succi02,Raabe07,Zhu05}. In contrast to those previous works we use a no-slip boundary condition which specifies a velocity is exactly on the lattice nodes and therefore is free of numerical slip. In Refs.\,\cite{Succi02,Zhu05} it is remarked that \eq{eq_slip_param} is basically a lattice implementation of a physical model of boundary events proposed by Maxwell back in the 19$^{th}$ century\,\cite{Maxwell1879,Maxwell1867}. 

The slip parameter is defined such that at a value of zero, the no-slip boundary condition is applied, while at a value of one, the full-slip boundary condition is applied. Thus, slip flows can be easily simulated by adjusting the value of $\zeta$ between zero and one according to physical parameters such as the roughness and hydrophobicity of the surface.

\paragraph{Specular Reflection boundary condition} 
The full-slip specular reflection boundary condition, as suggested in\,\cite{Succi01}, is used in the case of smooth boundaries with negligible resistance exerted upon the fluid, and thus the full-slip condition is obtained. Such a boundary condition implies no tangential momentum transfer to the wall.

In three dimensions, at the left ($x=0$) boundary, this reads as 
\[ f_{1} = f_{2}, 
\quad f_{8} = f_{12}, 
\quad f_{7} = f_{11}, 
\quad f_{9}  = f_{13}, 
\quad f_{10} = f_{14}\,. 
\]
At the right boundary we have 
\[ 
f_{2} = f_{1}, 
\quad f_{11} = f_{7}, 
\quad f_{12} = f_{8}, 
\quad f_{14} = f_{10}, 
\quad f_{13} = f_{9} \, . 
\]

\paragraph{Bounce back boundary condition} 
For the no-slip bounce back boundary condition, we use the formulation presented recently by Hecht and Harting\,\cite{Hecht2009}. This boundary condition is an explicit local on-site second order flux boundary condition for three dimensional LB simulations on a D3Q19 lattice.

When we combine this boundary condition with the specular reflection and insert them into \eq{eq_slip_param}, we obtain for the left ($x=0$) boundary, which we assume not to move in the x-direction:
\begin{eqnarray}
   \label{eq_bb_left_start}
   f_{1} &=& f_{2} \,,\\
   f_{8} &=& (1-\zeta) \cdot (f_{11} - \frac{\rho v_y}{6} + N^x_y) 
          + \zeta \cdot f_{12} \,,\\
   f_{7} &=& (1-\zeta) \cdot (f_{12} + \frac{\rho v_y}{6} - N^x_y) 
          + \zeta \cdot f_{11} \,,\\
   f_{9}  &=& (1-\zeta) \cdot (f_{14} + \frac{\rho v_z}{6} - N^x_z) 
          + \zeta \cdot f_{13} \,,\\
   f_{10} &=& (1-\zeta) \cdot (f_{13} - \frac{\rho v_z}{6} + N^x_z) 
          + \zeta \cdot f_{14} \,.
   \label{eq_bb_left_end}
\end{eqnarray}
with
\begin{eqnarray}
  N^x_y &=&  \frac{1}{2}\left[f_{3}+f_{15}+f_{16}-(f_{4}+f_{17}+f_{18})\right] 
      - \frac{1}{3}\rho v_y \label{eq_Nxy}\,,\\
  N^x_z &=&  \frac{1}{2}\left[f_{5}+f_{11}+f_{15}-(f_{6}+f_{16}+f_{18})\right] 
      - \frac{1}{3}\rho v_z \label{eq_Nxz}\,,
\end{eqnarray}
and
\begin{eqnarray}
    \rho &=& \left[ f_{3} + f_{4} + f_{5} + f_{6} \right.\nonumber\\
    &+&  f_{15} + f_{16} + f_{17} + f_{18} + f_{19} \label{eq_rhoinx_left} \\
    &+&\left.  2 ( f_{2} + f_{11} + f_{12} + f_{13} + f_{14})\right]\,.\nonumber
\end{eqnarray}

At the right boundary we have 
\begin{eqnarray}
   \label{eq_bb_right_start}
   f_{2} &=& f_{1} \,,\\
   f_{11} &=& (1-\zeta) \cdot(f_{8} + \frac{\rho v_y}{6} - N^x_y) 
           + \zeta \cdot f_{7} \,,\\
   f_{12} &=& (1-\zeta) \cdot(f_{7} - \frac{\rho v_y}{6} + N^x_y) 
           + \zeta \cdot f_{8} \,,\\
   f_{14} &=& (1-\zeta) \cdot(f_{9} - \frac{\rho v_z}{6} + N^x_z)  
           + \zeta \cdot f_{10} \,,\\
   f_{13} &=& (1-\zeta) \cdot(f_{10} + \frac{\rho v_z}{6} - N^x_z) 
           + \zeta \cdot f_{9} \,.
   \label{eq_bb_right_end}
\end{eqnarray}
with
\begin{eqnarray}
    \rho &=& \left[ f_{3} + f_{4} + f_{5} + f_{6} \right.\nonumber\\
    &+&  f_{15} + f_{16} + f_{17} + f_{18} + f_{19} \label{eq_rhoinx_right} \\
    &+&\left.  2 ( f_{1} + f_{7} + f_{8} + f_{9} + f_{10})\right]\,.\nonumber
\end{eqnarray}                                      

The values for $N_y^x$ and $N_z^x$ are according to \eqns{eq_Nxy}{eq_Nxz}.

Note that for the specular reflection boundary condition, the correction terms proposed by Hecht and Harting are not used\,\cite{Hecht2009}. Instead we stick to the original description of specular reflection as mentioned in\,\cite{Succi01}, applying it for three dimensions, so that tangential momentum is conserved. Results show that a second order accuracy is however maintained, when the linear combination of the two boundary conditions is used to model slip flows: the parabolic flow profile is obtained exactly without kinks or jumps close to the boundary, as shown in \fig{fig:profile}.

\section{Results and Discussion}

\paragraph*{Simulation Setup:} The simulation volume consists of a cubic box of 32 nodes in the $x$, $y$, and $z$-directions, unless stated otherwise. Periodic boundary conditions are set up along the $y$ and $z$-directions. The boundary conditions given in the previous section are applied in $x$-direction. A constant accelerating force is applied to the whole domain. The $x$-component of the force is always zero, but the force can be turned continuously in the $yz$-plane. Usually, the force acts along the $z$-direction, except in the section where we explicitly turn the force. The setup is relaxed for 10,000 to 20,000 lattice updates to attain equilibrium. The profile of fluid flow between two plates is studied in three dimensions for varying slip parameter $\zeta$, varying single relaxation times, varying force or shear velocity. Unless stated elsewise we use $\tau=1$ throughout the paper.

\subsection{Homogeneous Walls}
We conduct various simulations for the slip parameter $\zeta$ and find that the slip parameter gives rise to a slip length which is independent of the shear rate. The results are consistent with the findings by Kunert\etal\,\cite{harting-kunert-herrmann-06}, who apply a Shan-Chen force\,\cite{shan-chen-93,shan-chen-liq-gas}. In our simulations, the slip length, measured in lattice units and obtained at a given value of the slip parameter, is found to be independent of the channel width. In contrast to slip models based on a Shan-Chen force\,\cite{harting-kunert-herrmann-06}, the slip length obtained in our model is found to be independent of the density of the fluid. However, if a density or pressure dependence is desired, the slip parameter can be set up as a function of the local density $\rho(\bf{x},t)$ in the current time step.

\begin{figure}[t]
\centering
\includegraphics[width=\figwidth]{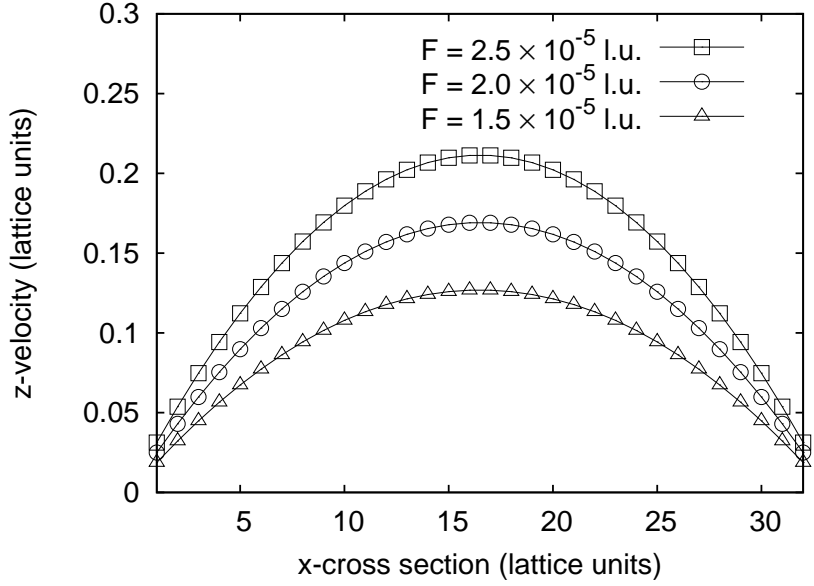}
\caption{A perfect parabolic profile at different accelerating forces. The partial slip boundary condition is applied at the nodes 1 and 32 respectively. The slip length is independent of the applied force, even when the force is varied over several orders of magnitude (not shown).}
\label{fig:profile}
\end{figure}

\begin{figure}[h]
\centering
\includegraphics[width=\figwidth]{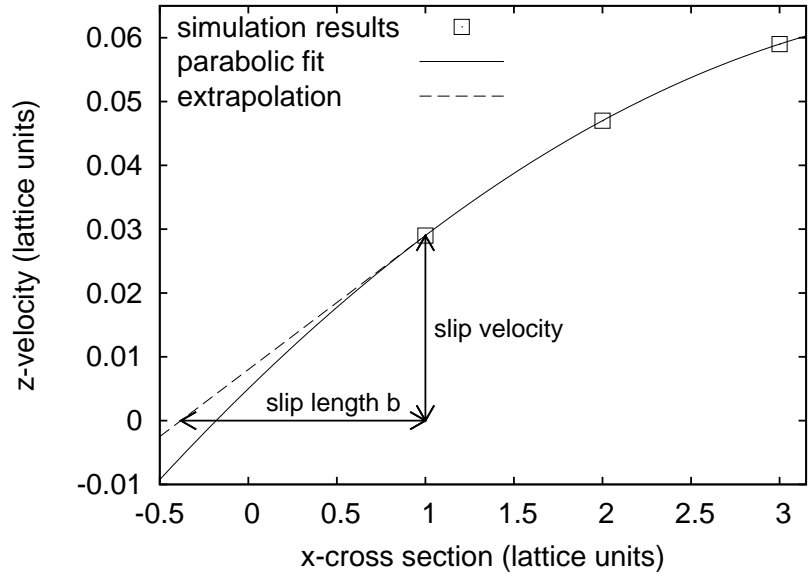}
\caption{Calculation of slip length at the wall boundary node: the boundary condition is applied at $x=1$. The extrapolated velocity profile thus equals zero at $ x = 1-b $.
For this schematic sketch simulation data of a channel with a width of 8 nodes was used.}
\label{fig:slip_lengthcal}
\end{figure}

The simulations give rise to a perfect parabolic profile for different values of the acceleration, as can be seen in \fig{fig:profile}. The accelerating force is varied over an order of a few hundreds, but this does not affect the slip length calculated in all cases, when the slip parameter is kept constant. For each simulation, the slip length of the flow profile is calculated by the linear extrapolation of the parabolic fit function, as shown in \fig{fig:slip_lengthcal}.  

\begin{figure}[t]
\centering
\includegraphics[width=\figwidth]{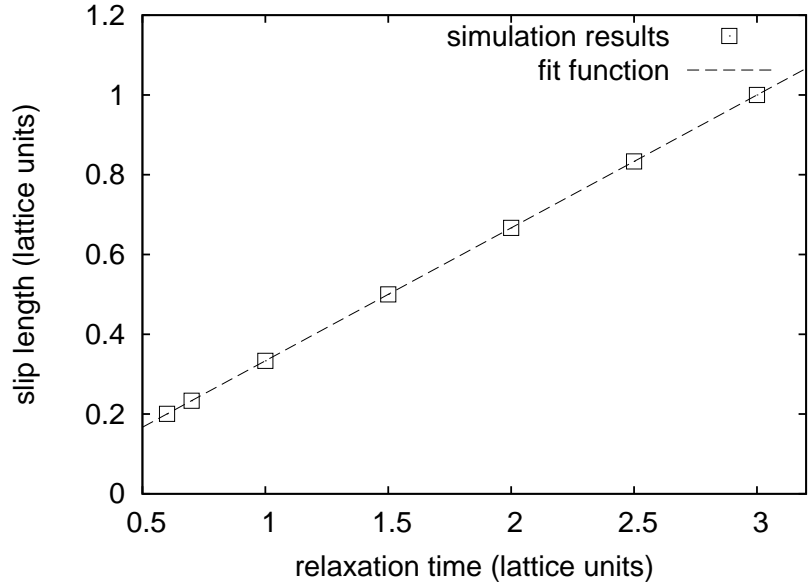}
\caption{Slip length  $b$ plotted against the relaxation time $\tau$ of the BGK collision operator, where the slip parameter $\zeta=0.5$ is kept constant. }
\label{fig:slip_tau}
\end{figure}

The slip length is found to vary in a linear manner only with the relaxation time of the simulation as shown in \fig{fig:slip_tau}. This dependency is important if the lattice
Boltzmann parameters are mapped to physical units, e.g. when aiming to adjust a specific Knudsen number in the simulation. Schemes how to map the simulation parameters to physical
quantities have been shown in the literature\cite{Ansumali07, Ansumali08, Guo08, Guo07}.
Once the relaxation time is chosen, the slip parameter may can be calculated according to \eq{eq_sliplen}.

The slip length in our simulations does not depend on the shear rate. This is consistent with many experiments. However, if dissolved gas forms bubbles 
at the surface, this may lead to to a shear rate dependence as shown in simulations\,\cite{Jari08}. The presence of gas bubbles might be a possible explanation for the shear rate dependence reported from some experiments\,\cite{lauga-brenner-stone}.

\begin{figure}[t]
\centering
\includegraphics[width=\figwidth]{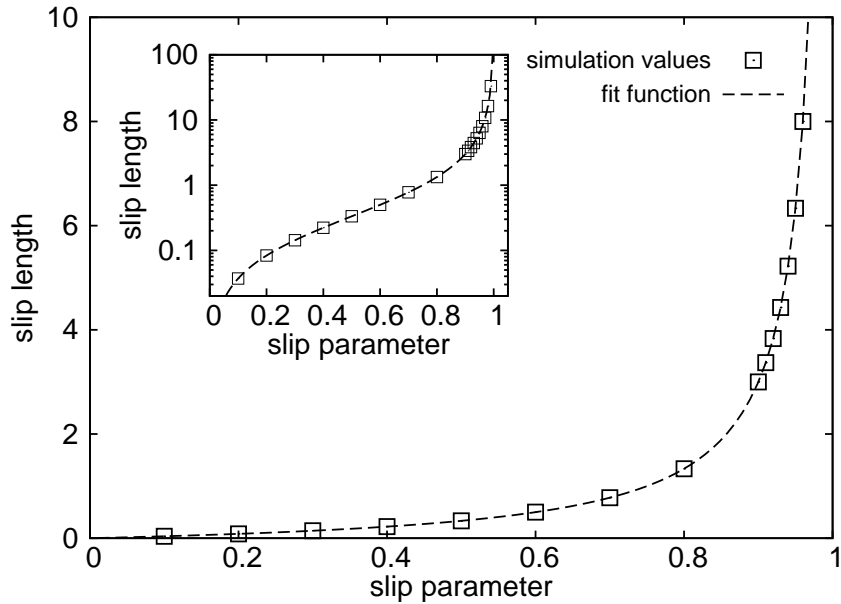}
\caption{Slip length $b$ plotted against the slip parameter $\zeta$. The inset contains the semi-log plot of the slip length against the slip parameter. The simulations for $\zeta$ $\epsilon$ $\lbrack$ 0.9,1$\rbrack$ are conducted in an enlarged system of $64 \times 32 \times 32$ lattice nodes allowing a more accurate fit of the parabolic profile at large slip lengths.}
\label{fig:slip_velocity}
\end{figure}

The slip length in our simulations can be tuned by the slip parameter $\zeta$. The slip length
must diverge for $\zeta \rightarrow 1$ and it approaches zero for $\zeta \rightarrow 0$.
Therefore we choose a power law of the form $b \sim 1/(\zeta^{-\alpha}-1)$ and determine the prefactor and 
the exponent from the simulation data. When fitting this phenomenological formula to the numerical data, it turns out that the exponent of $1$ and a prefactor of $1/3$ fits best. The relation therefore simplifies to the following form:
\begin{equation}
\frac{b}{a} = \frac{\tau\,\zeta}{3\,(1 - \zeta)}\,, \label{eq_sliplen}
\end{equation}
where $b$ is the slip length, $a$ is the lattice constant, $\tau$ is the relaxation time in the LBGK dynamics, and $\zeta$ is the slip parameter defined according to \eq{eq_slip_param}. \eq{eq_sliplen} matches the numerical results with a relative error of less than $0.03\%$, 
or $10^{-3}$ lattice units, on the entire range of $\zeta$, where we can obtain slip lengths 
ranging from fractions of a lattice unit up to more than 30 lattice units. In 
\fig{fig:slip_velocity} both, the numerical results and \eq{eq_sliplen} are plotted.
At very large slip lengths the correspondence to the analytical Boltzmann equation 
may be questionable. Our approach is purely phenomenological and if the mean free path
becomes comparable to the system size a more accurate treatment of the Knudsen layer 
is advised. In Ref.\,\cite{Succi02} a similar approach as ours is followed, but for the 
no-slip part the correction terms $N^i_j$ are not applied. Therefore, \eq{eq_sliplen}
cannot be applied directly to estimate the slip in these simulations. 

\subsection{Homogeneous Moving Walls}
\begin{figure}[t]
\centering
\includegraphics[width=\figwidth]{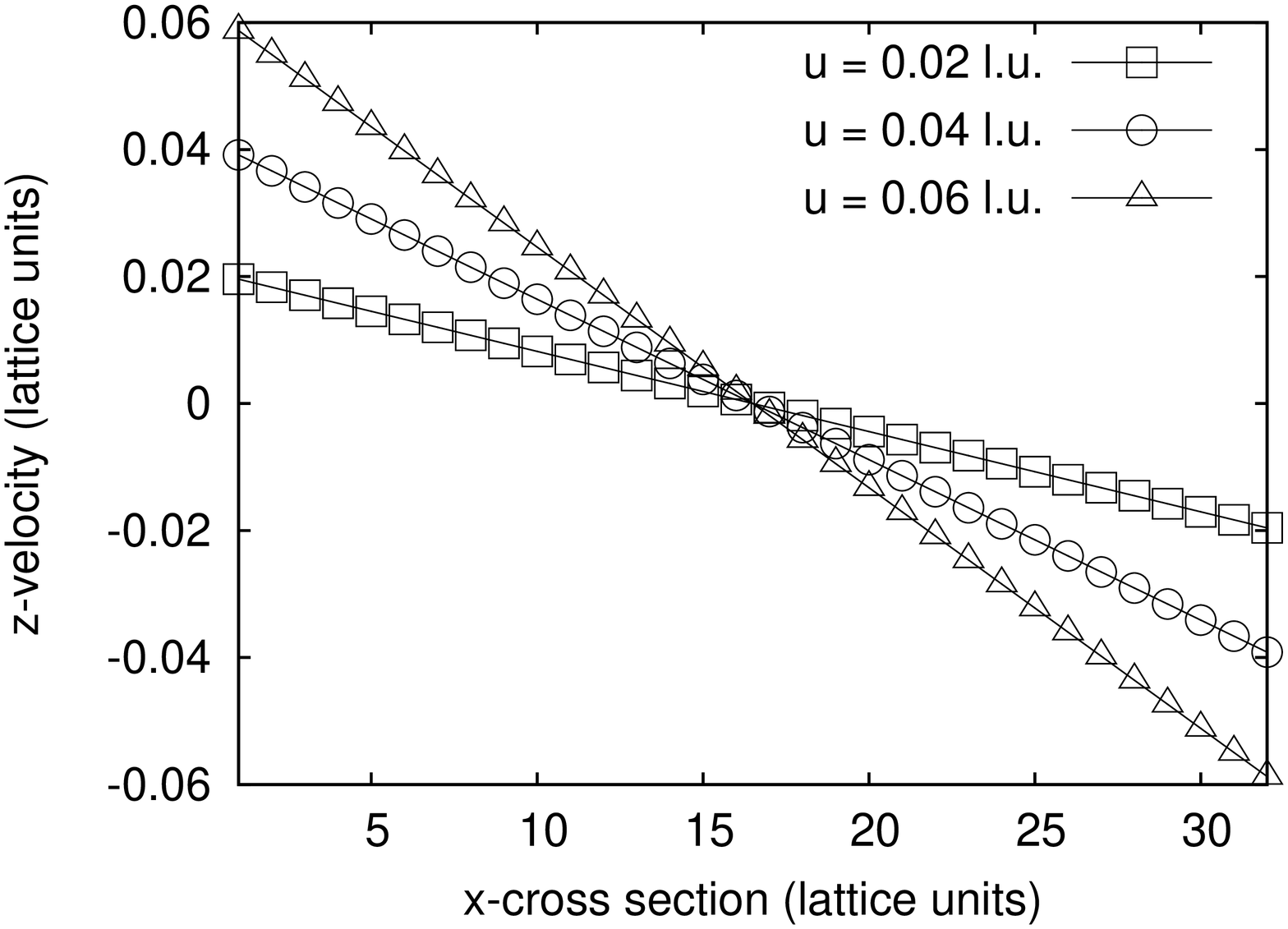}
\caption{Flow profiles for shear flow at different wall velocities. The slip parameter $\zeta$ is set to $0.5$ for these simulations. The velocity at the wall does not reach exactly the one at which 
the wall is moving, but a non-vanishing slip length consistent with the results found for Poiseuille flow can be extracted from the numerical data. This shows that the boundary condition also works properly for moving walls.}
\label{fig:moving_wall}
\end{figure}

The boundary conditions are also tested for their applicability in different flow profiles by considering a moving wall. A shear velocity in $z$-direction is added to the boundary nodes giving rise to a velocity gradient along the $x$-direction, and the slip behaviour is studied, as shown in \fig{fig:moving_wall}. The accelerating forces are set to zero in these shear simulations. The results for the slip length match with the findings in the case of a constant accelerating force on the domain. This confirms that these combined boundary conditions can be applied to successfully simulate fluid-wall interaction causing the fluid to move.

From the above simulations, we understand that on using the combined boundary conditions, we get a slip length $b$ that is dependent only on the slip parameter $\zeta$ and is not affected by parameters such as the shear rate and channel width. The simulations maintain a second order numerical accuracy as we do not see any errors in the parabolic fit even along the wall boundary nodes.

\subsection{Textured Walls with Stripes}
We also study the application of the boundary conditions to modified walls that carry a 
texture of thin stripes of varying slip parameter. We focus on stripes which have equal 
widths and alternating slip parameter $\zeta_1$ and $\zeta_2=1-\zeta_1$. To enhance the 
influence of the structured wall on the flow behavior we use a simulation box of 
$16\times 32 \times 32$ lattice nodes, where the shorter extension is the thickness of
the channel, and the two larger extensions are the directions in the planes of the 
walls.

The effective slip length of the flow through such a channel is calculated. The slip 
length is found to be a second order function of the slip length obtained from homogeneous walls, given by the form:

\begin{equation}
b_{m} = A \cdot (b_{1} + b_{2}) + B \cdot (b_{1}^2 + b_{2}^2) + C
\label{slip_mod}
\end{equation}

where $b_m$ is the slip length of the modified wall with stripes, $b_{1}$ and $b_{2}$ are the slip lengths obtained for the homogeneous walls of the same slip parameters $\zeta_1$ and $\zeta_2$, 
respectively.  $A$, $B$, and $C$ are prefactors, which do not depend on the slip parameters. 
However, they still depend on the orientation of the stripes on the wall and on the width of 
the stripes of which the pattern is formed. 

\begin{figure*}[h]
\centering
\begin{minipage}{\linewidth}
\mbox{a)\includegraphics[width=0.47\linewidth]{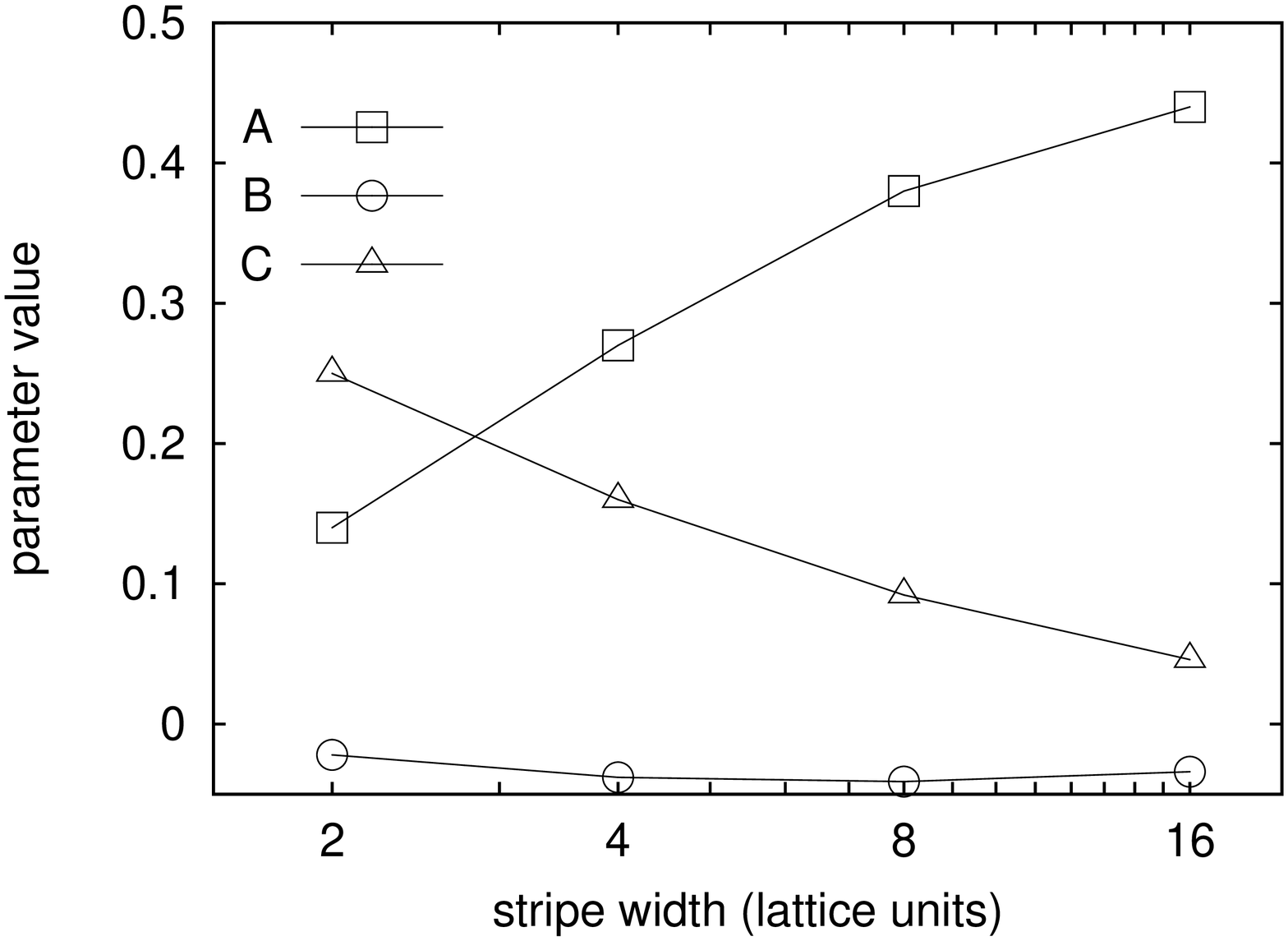}
b)\includegraphics[width=0.47\linewidth]{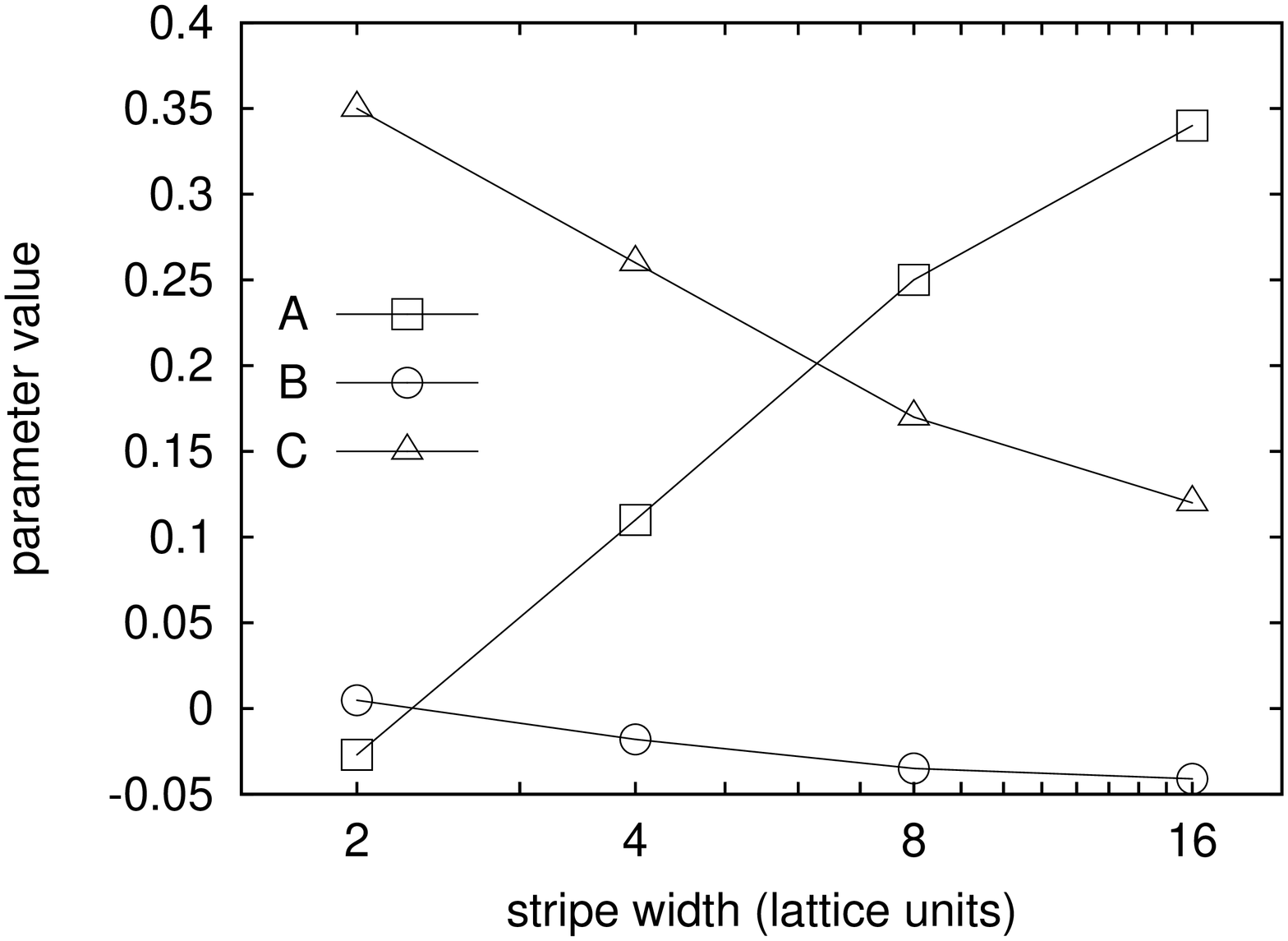}}
\end{minipage}
\caption{Parameters $A, B,$ and $C$ in \eq{slip_mod} depending on the width of the stripes 
for the principal orientations of the stripes: a) parallel, and b) perpendicular to the 
direction of the accelerating force.}
\label{fig:params}
\end{figure*}

For the fit parameters $A$, $B$, and $C$ in \eq{slip_mod}, depending on the width of the stripes 
and their orientation, we find the values listed in \tab{tab:parameters}. The dependence of 
the parameters on the width of the stripes is plotted in \fig{fig:params} for the principal 
cases of the stripes oriented parallel to the force (a) and perpendicular to the force (b), 
respectively.

\begin{table*}
\begin{minipage}{\linewidth}
\mbox{
\begin{tabular}{|l||r|r|r|r|}
\hline
width (l.u.) &$ 2 $&$ 4 $&$ 8 $&$ 16 $\\\hline\hline
$A$ &$ -0.027 $&$ 0.11 $&$ 0.25 $&$ 0.34 $\\\hline
$B$ &$ 0.0048 $&$ -0.018 $&$ -0.035 $&$ -0.041 $\\\hline
$C$ &$ 0.35 $&$  0.26 $&$ 0.17 $&$ 0.12 $\\\hline
\end{tabular}
\quad
\begin{tabular}{|l||r|r|r|r|}
\hline
width (l.u.) &$ 2 $&$ 4 $&$ 8 $&$ 16 $\\\hline\hline
$A $&$ 0.14 $&$ 0.27 $&$ 0.38 $&$ 0.44 $\\\hline
$B $&$ -0.022 $&$ -0.038 $&$ -0.041 $&$ -0.034 $\\\hline
$C $&$ 0.25 $&$ 0.16 $&$ 0.092 $&$ 0.046  $\\\hline
\end{tabular}}
\end{minipage}
\caption{Fit parameters $A$, $B$, and $C$ in \eq{slip_mod} in the case of 
the stripes oriented perpendicular to the flow(left), and parallel to the flow (right), 
depending on the width of the stripes.}
\label{tab:parameters}
\end{table*}

The accelerating force is always applied in the whole domain and is oriented parallel 
to the walls. Consistently with the work by Feuillebois\etal\,\cite{Feuillebois}, we find 
that the maximum slip length is obtained when the stripes are aligned parallel to the 
force. However, the orientation of the force can be continuously varied between the 
principal cases considered up to now. 

\begin{figure}[h]
\centering
\includegraphics[width=\figwidth]{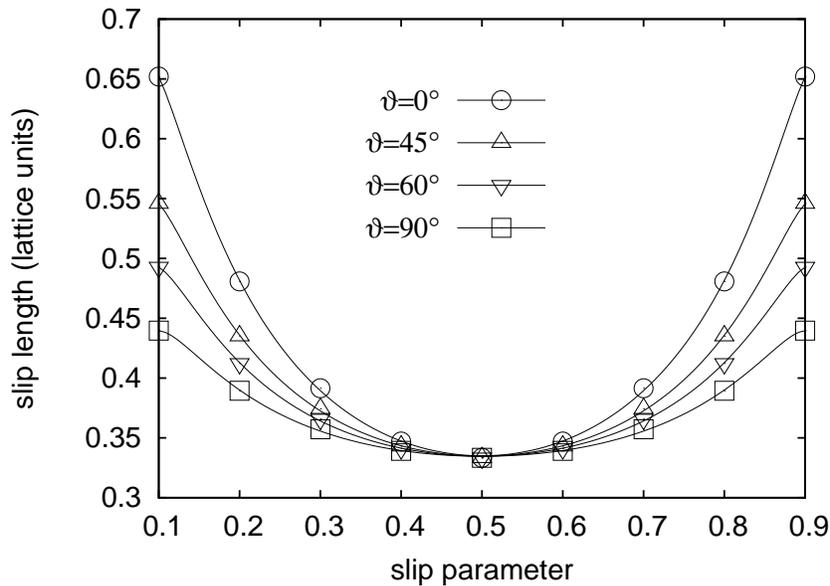}
\caption{Slip length of a striped wall as a function of the slip parameter of the stripes 
for four different orientations of the stripes with respect to the direction of the 
driving force: parallel ($\theta = 0^\circ$), perpendicular ($\theta = 90^\circ$), 
diagonal ($\theta = 45^\circ$), and tilted by 60 degrees. The width of the stripes is set 
to 4 lattice sites for all cases in this plot. The slip lengths measured in the simulations
for the diagonal and tilted case agrees up to a relative error of $10^{-3}\%$ with the 
theoretical prediction in \eq{eq:bazant}.}
\label{fig:sliplength_diag_fit}
\end{figure}

The dependence of the slip length for walls with stripes oriented at the angle $\theta$ between the 
direction of the force and the extension of the stripes has been 
previously calculated analytically by Bazant and Vinogradova\,\cite{Bazant}, to be
\begin{equation}
  b = b_\parallel \cos^2\theta + b_\perp \sin^2 \theta\,.\label{eq:bazant}
\end{equation}
Here we just consider the component of the slip parallel to the driving force.
We find that the slip length obtained in our simulations using the tilted force agrees with 
the theoretical prediction of \eq{eq:bazant} with a relative error of only $10^{-3}\%$. 
In \fig{fig:sliplength_diag_fit} we show the dependence of the effective slip length
on the orientation of the force and on the slip parameter applied on the stripes. 
For $\zeta_1 = \zeta_2 = 0.5$ the case of a homogeneous wall is restored and therefore, 
the slip length is independent on the angle $\theta$ between the force and the direction 
of the extension of the stripes. For $\theta=0^\circ$, the force is aligned parallel to 
the stripes, and therefore, a maximal slip length is obtained for this case. The slip length
increases with the contrast between the slip parameters for the individual stripes of high
and low slip parameters $\zeta_1$ and $\zeta_2$.

We also find that in simulations with the force acting in diagonal direction, 
the fluid tends to follow the stripes on the wall. The velocity of the fluid shows
a small component which is perpendicular to the force. This flux is four orders of magnitude 
less than the velocity obtained in the direction of the force. Even though the effect is small, 
these simulations confirm that walls with stripes of varying slip length can be used in mixing 
devices to cause vortices in micro and nano flows\,\cite{Stroock04}.

\begin{figure*}[h]
\centering
\begin{minipage}{\linewidth}
\mbox{a)\includegraphics[width=0.47\linewidth]{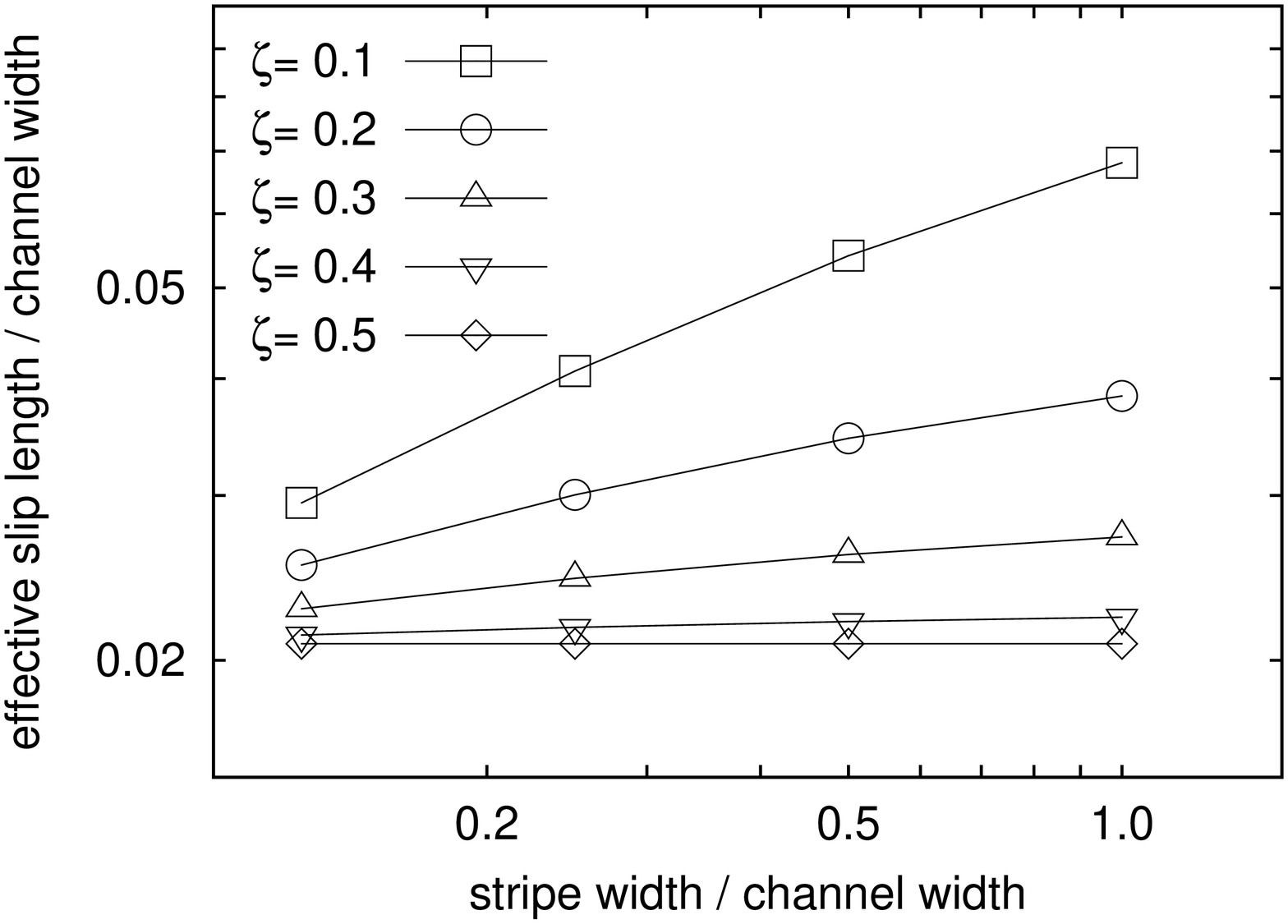}
b)\includegraphics[width=0.47\linewidth]{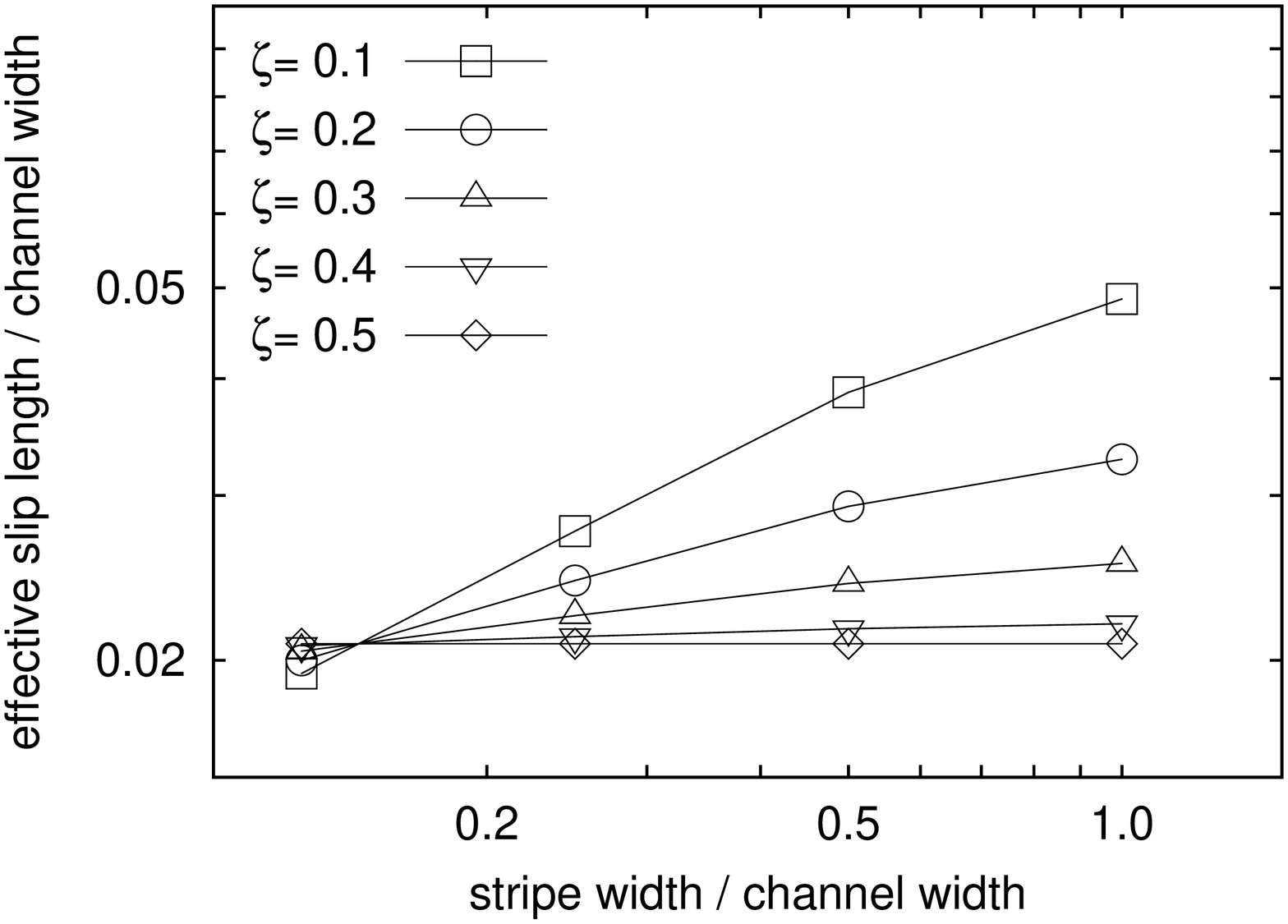}}
\end{minipage}
\caption{Dependence of the effective slip length on the width of the stripes for different
slip parameters for stripes parallel (a) and perpendicular (b) to the driving force.}
\label{fig:stripe_width}
\end{figure*}

Simulations are also conducted to study the relation between the slip length obtained and the width 
of the stripes for a given slip parameter. Similar to the work by Priezjev\etal\,\cite{Priezjev05}, 
who studied the dependence of the slip length on the width of the stripes in patterns of 
alternating full- and noslip-stripes, we find a power-law behavior of the slip length
(\fig{fig:stripe_width}) as long as the width of the stripes is small compared to the 
channel width, i.e., in the limit of thin stripes. For thicker stripes the slip length starts to 
saturate. The dependence of the effective slip length on the width of the stripes is more pronounced 
the larger the contrast between the slip parameters. For $\zeta_1 = \zeta_2 = 0.5$, i.e., in the 
case of a homogeneous wall, the effective slip length stays constant. 

These observations reflect what one would expect qualitatively. The authors are not
aware of a theory which would give quantitative results which the simulation data
could be compared to.

\begin{figure*}[h]
\centering
\begin{minipage}{\linewidth}
\mbox{a)\includegraphics[width=0.47\linewidth]{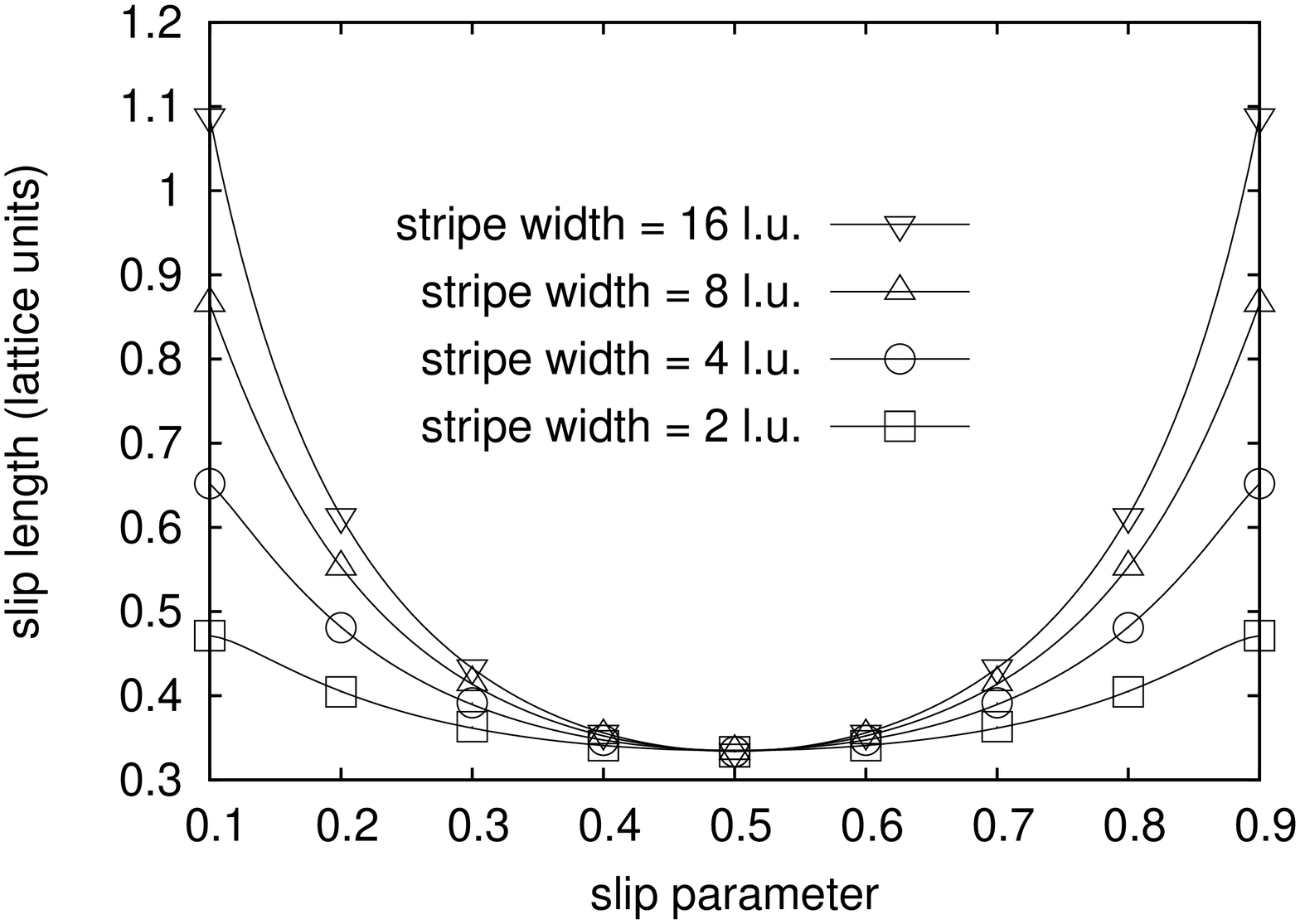}
b)\includegraphics[width=0.47\linewidth]{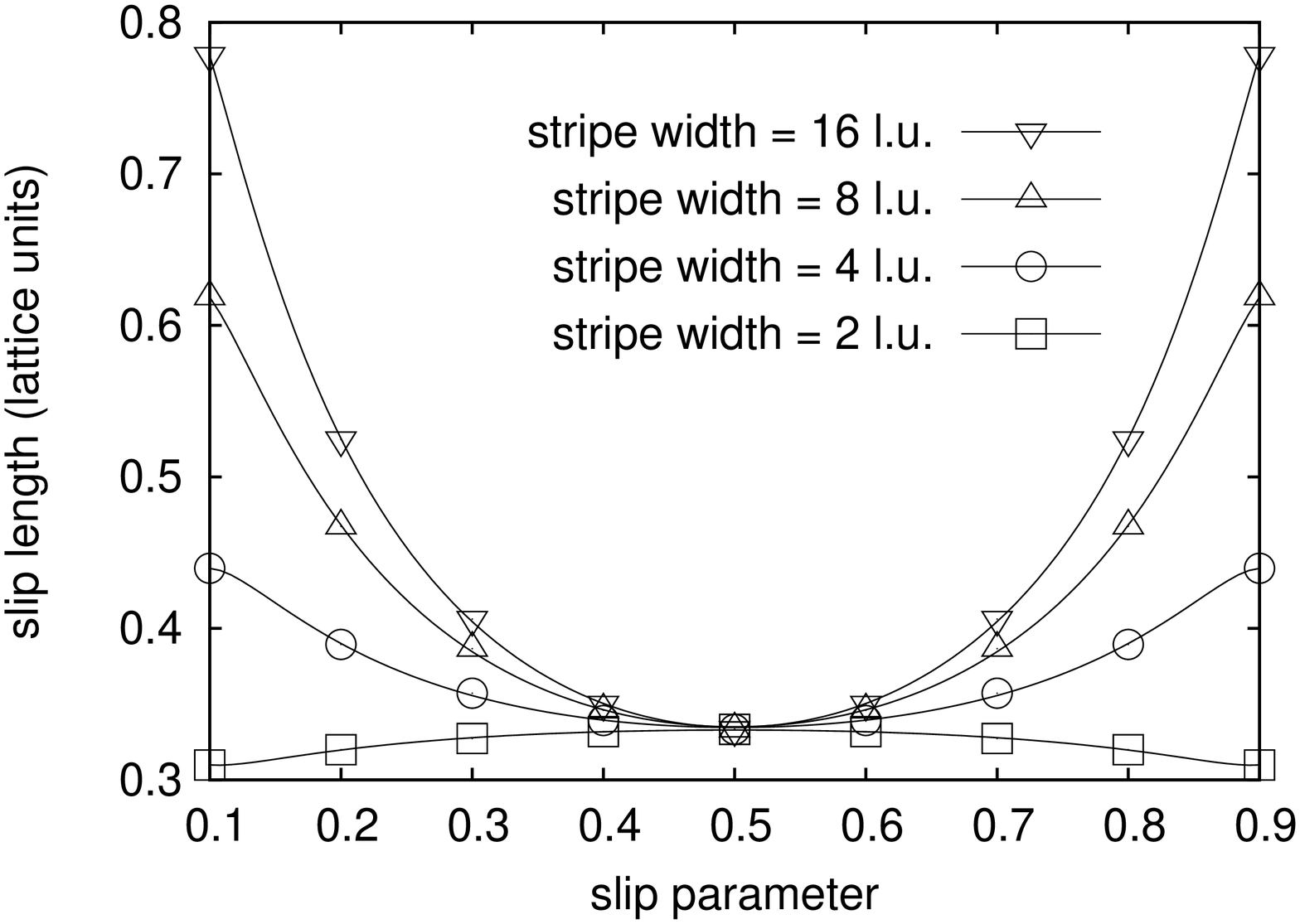}
}
\end{minipage}
\caption{Slip length plotted against the slip parameter for stripes of different widths, with the orientation of the stripes a) parallel and b) perpendicular to the direction of the accelerating force.}
\label{fig:width_dependence}
\end{figure*}

If we compare the different patterns for varying slip parameters, we find again, that the 
simulation data can be fitted by a polynomial expression \eq{slip_mod} for the slip lengths 
corresponding to homogeneous surfaces with the slip parameter $\zeta_1$ and $\zeta_2$, 
respectively. In \fig{fig:width_dependence} the fits are shown together with the simulation 
results. The parameters obtained for the fits are listed in \tab{tab:parameters}.
For a fixed width of the stripes these parameters can be used to predict the slip length at a 
given slip parameter $\zeta_1$ in the direction parallel as well as perpendicular to the stripes.
Then, \eq{eq:bazant} can be used to calculate the slip length at arbitrary angle $\theta$ 
of the flow direction with respect to the orientation of the stripes. 

Note that in this paper we have restricted ourselves to patterns consisting of stripes of 
equal width and slip parameters which fulfill the condition $\zeta_2 = 1 - \zeta_1$. If 
either of these two restrictions is released, the fit function \eq{slip_mod} needs to be
modified due to the breaking of the symmetry. The situation becomes much more complex and 
therefore this more general case is left for future investigations which will be published 
elsewhere.

\section{Conclusion}
The above results show that the proposed combined boundary conditions can be used to model slip flows accurately. The slip parameter of this model allows to adjust a given slip length, which either has to be determined experimentally for a given surface or can be obtained by carrying out molecular dynamic simulations which relate the slip length to specific fluid-surface interactions. The slip length in our model is found to be independent of the density, proportional to the lattice constant, and linearly dependent on the BGK relaxation time. With this model, slip lengths of several tens of lattice units can be obtained.
Hence by using these boundary conditions, once a study of the slip length of a material is conducted, the behaviour of the fluid flow along such materials can be simulated using LBM.
For surfaces patterned with stripes of alternating large and small slip length we 
have shown that the effective slip length can be calculated from the slip lengths of 
the individual regions using a polynomial expression. The parameters depend on the width of 
the stripes and on their orientation. For the orientation perfect agreement with the theoretical
prediction is found, whereas for the dependence on the stripe width qualitative agreement 
with similar numerical data from the literature is given, but an exact theory is still to develop.

\section{Acknowledgment}
N.K.A. thanks the German Academic Exchange Service (DAAD) for a scholarship under the Working Internships in Science, Technology and Engineering (WISE) scheme for the summer of 2009. M.H. thanks the German Research Foundation (DFG) for financial support within grant EAMatWerk. The authors thank Jens Harting for fruitful discussions.

\end{document}